# Temperature dependence of a nanomechanical switch

Robert L. Badzey[a)], Guiti Zolfagharkhani[a)], Alexei Gaidarzhy[b)] & Pritiraj Mohanty[a)]

a) Department of Physics, Boston University, Boston, Massachusetts 02215

b) Department of Aerospace and Mechanical Engineering, Boston University, Boston, Massachusetts 02215

We present the effect of temperature on the switching characteristics of a bistable nonlinear nanomechanical beam. At MHz-range frequencies, we find that it is possible to controllably change the state of the system between two stable mechanical states defined by the hysteresis brought on by nonlinear excitation. We find that the introduction of increased temperature results in a loss of switching fidelity, and that temperature acts as an effective source of external noise on the dynamics of the system.

The realization of mechanical memory and processing elements has been a goal ever since the design of the first mechanical analytic engine by Babbage in 1834[1]. Recent work[2] has demonstrated the feasibility of using a nanomechanically-fabricated doubly-clamped silicon beam as a memory element. By exploiting the bistability brought about by nonlinear excitation, this element has demonstrated binary, controllable switching when subjected to an external modulation. Reducing the modulation power results in a loss of switching fidelity, but the size of the switch remains consistent.

We have fabricated a doubly-clamped beam structure from single-crystal silicon by electron-beam lithography and nanomachining as the basis for our nanomechanical memory element. This beam has a length of 8 μm, a width of 300 nm and a thickness of 200 nm, with a corresponding resonance frequency in linear response of 23.568 MHz. The dynamics of this system have been well-understood since the time of Euler and have been the subject of numerous studies[3,4,5]. For small driving forces, the beam response is linear and follows a Lorentzian lineshape in frequency space. However, as the driving force is increased, the strain becomes plastic and the beam length increases. Because the beam is clamped between two large pads which are a fixed distance apart, the lengthened beam experiences an additional compressive strain due to this geometric restriction. There are actually two stable points of oscillation, and the beam is governed by the well-known Duffing equation

$$\ddot{x} + 2\gamma\dot{x} + \omega_0^2 x \pm k_3 x^3 = F\cos\omega t. \quad (1)$$

At this point, the beam response is bistable and can oscillate linearly about the two fixed points described by a double-well potential. Sweeping the frequency up and down shows a marked hysteresis; this defines a region in which the solution of the beam equation is multi-valued[6]. The frequency sweep shows a characteristic sharp drop, which appears on either the left-hand side or right-hand side, depending on whether the beam softens (left) or hardens (right).[7]

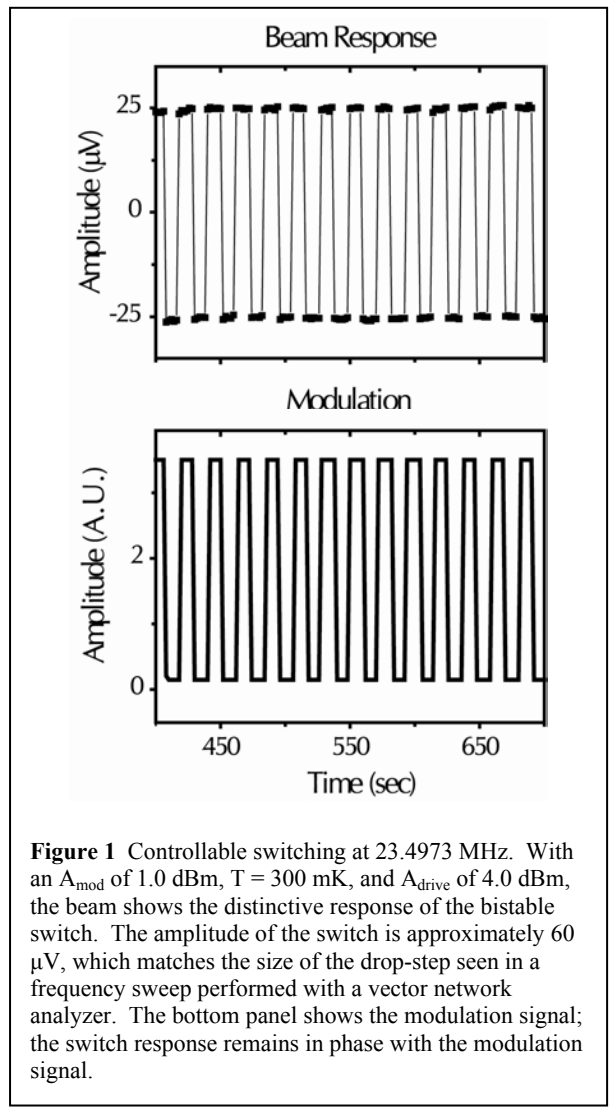

**Figure 1** Controllable switching at 23.4973 MHz. With an $A_{mod}$ of 1.0 dBm, T = 300 mK, and $A_{drive}$ of 4.0 dBm, the beam shows the distinctive response of the bistable switch. The amplitude of the switch is approximately 60 μV, which matches the size of the drop-step seen in a frequency sweep performed with a vector network analyzer. The bottom panel shows the modulation signal; the switch response remains in phase with the modulation signal.

We have taken advantage of this characteristic of the beam by driving our nanomechanical beam with a large amplitude force at a single frequency within this bistable region. We then add an external modulation which can controllably switch the beam between its two stable states. The beam is driven through the use of the well-known magnetomotive technique[2,8,9,10] at 300 mK. Figure 1 shows the switch response with a driving frequency of 23.497 MHz, $A_{drive}$ = 4.0 dBm, $A_{mod}$ = 1.0 dBm, and $f_{mod}$ = 0.05 Hz.

An important characteristic of this hysteretic region is the temperature dependence. In general, an increase in temperature results in a decrease in the region width and an overall shift to higher frequencies. Figure 2 shows the evolution of the hysteresis with increasing temperature. In addition, the hysteretic region is not the perfect square shape one would assume from theory; rather, it shows a finite slope on both the upper state and lower state branches. The upper and lower states also show more structure than a simple flat line, with areas where the gap size decreases or increases depending on the driving frequency.

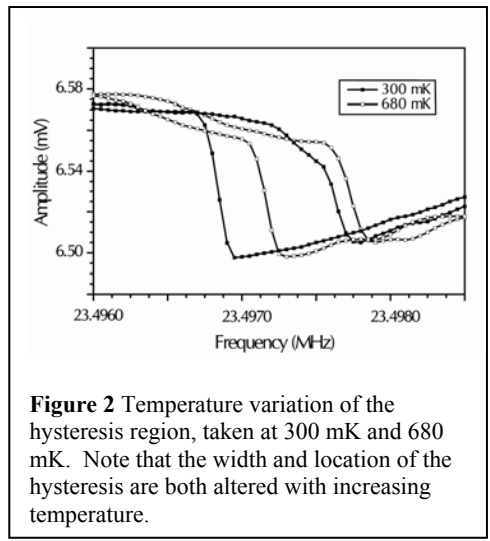

**Figure 2** Temperature variation of the hysteresis region, taken at 300 mK and 680 mK. Note that the width and location of the hysteresis are both altered with increasing temperature.

Understanding the higher-temperature characteristics of our switch is vital, as any realistic implementation of a nanomechanical memory element demands that the effects of higher temperature be explored. Effects on the switching behavior will have their roots, at least in part, in the temperature dependence on the shape and size of the bistable region itself. Although these changes are slight, they do reflect the necessity of ensuring that the driving frequency is well within the bistable region, so that the temperature-induced changes to the bistable region have minimal impact. It is for this reason that we have used a driving frequency of 23.4973 MHz, which is well within the bistable region at all relevant temperatures. Therefore we can reasonably deduce that any effects seen are due to the interaction between the temperature bath and the dynamics of the beam, and have little dependence on the evolution of the hysteresis region itself.

Having established that the beam is being driven at a frequency which should allow for bistable switching, we explore the effects of increasing temperature on the switching itself. Figure 3a shows the effect of increased temperature on the switching characteristics of the beam. The first panel depicts full switching consistent with 1.0 dBm of modulation power at 300 mK, with a histogram plot showing the splitting of the signal into two separate and distinct states. The second panel depicts the beam response at 825 mK. There are three major effects that are brought on by the increase in temperature. First, the spread of the points increases, reflecting the increasing noise resulting from a higher temperature. This is easily seen from the widening of the histograms for each state. Secondly, the size of the gap decreases with increasing temperature. Part of this may be attributed to the fact that we have not changed the driving frequency, and are therefore accessing a part of the bistable region which has a slightly smaller gap. Finally, the switching fidelity decreases; several periods of switches are skipped within any scan. Even though switches are skipped, the phase of the switching is maintained. After a skipped switch, the bridge will not switch again until at least one period has passed. Sometimes several periods will pass before the bridge switches again, but regardless of the amount of time passed, the bridge will still only switch in phase with the modulation. Figure 3b includes the data from all temperatures studied here, showing the overall downward trend in gap size and increase in the noise the signal (the FWHM of each histogram, added in quadrature), with increasing temperatures. Figure 3c shows the increase of skipped periods with increasing temperature; the residence time fraction is the time spent in either the upper or lower state. At full switching, the residence time fraction for each state would therefore be 0.5.

In conclusion, we find that increasing the ambient temperature of the beam results in a loss of switching fidelity and general degradation of the character of the switch. The temperature bath acts as a well-coupled source of external noise which profoundly affects both switching fidelity and the overall character of the nonlinear bistable states. However, there is a possibility that these temperature effects need not be fatal to the overall utility of this bistable beam system. Of particular importance is the phenomenon of stochastic resonance[11] which involves the counter-intuitive notion that under certain conditions, adding noise to a modulated bistable system can actually increase its coherent response. Having established that the temperature bath is a viable method of coupling external noise to the beam, it is possible that at the right temperatures, one would actually see an increase in the coherent response of the beam switching, rather than the expected decrease. Current and ongoing investigations of this system are exploring this very possibility.

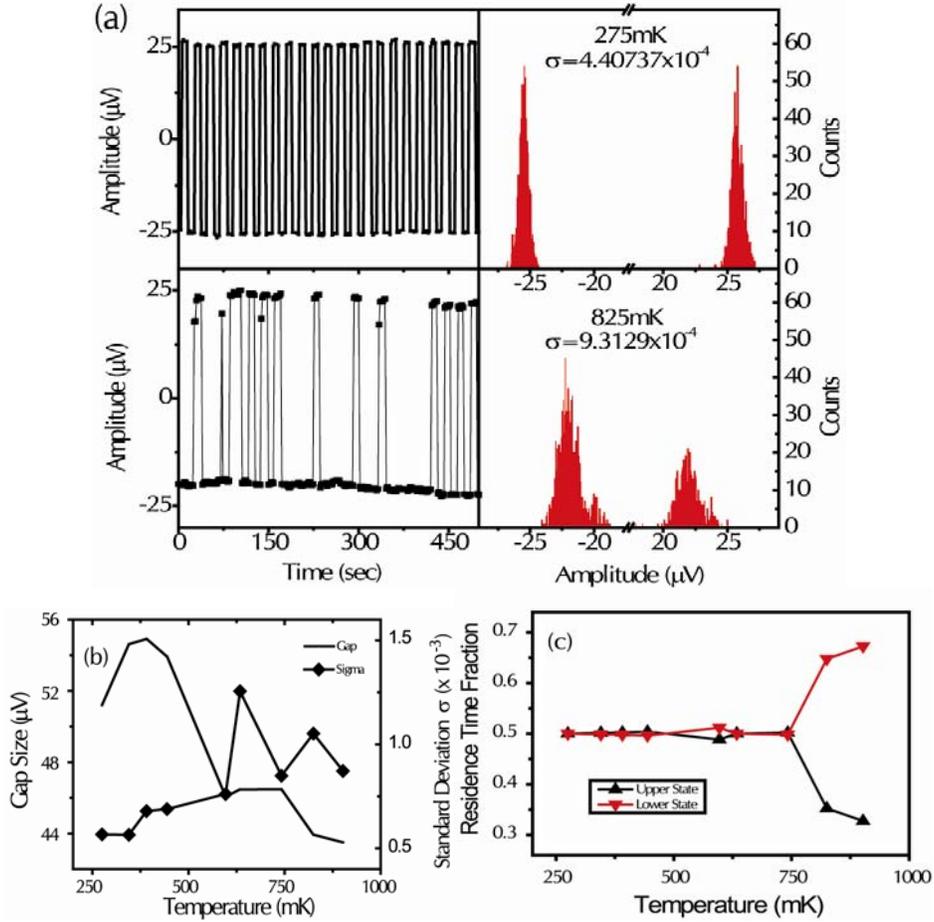

**Figure 3** Effect of temperature on switching control and magnitude. (a) Representative scans at 300 mK and 825 mK, with associated histogram plots depicting the separation of the signal into two separate and distinct states. At low temperature, the switching is full and complete; the histogram shows two sharp peaks centered at the mean of each state. As temperature is increased, the noise of each state also increases. Switching fidelity is also lost, as can be seen from the definite disparity between the sizes of the respective histogram peaks. (b) Graph of the changes in gap size (solid line) and the average state noise as defined by the width of each histogram peak (diamonds) with temperature. Increases in temperature lead to an overall decrease in gap size, with an overall increase in state noise. (c) Plot of residence time fraction (RTF) for each state as a function of temperature. As the temperature is increased the RTF for the lower state (downward pointing triangles) increases, while that of the upper state (upward pointing triangles) decreases correspondingly.

We acknowledge the NSF (Nanoscale Exploratory Research (NER) grant number ECS-0404206), and the DOD/ARL (DAAD 19-00-2-0004). We also acknowledge partial support from the Sloan Foundation and the NSF (DMR-0346707, CCF-0432089 and ECS-0210752).